# Concept of spectrometer for resonant inelastic X-ray scattering with parallel detection in incoming and outgoing photon energies

V.N. Strocov

Swiss Light Source, Paul Scherer Institute, 5232 Villigen-PSI, Switzerland

A spectrometer for resonant inelastic X-ray scattering (RIXS) is proposed where imaging and dispersion actions in two orthogonal planes are combined to deliver full two-dimensional map of RIXS intensity in one shot with parallel detection in incoming $hv_{in}$ and outgoing $hv_{out}$ photon energies. Preliminary ray-tracing simulations with a typical undulator beamline demonstrate a resolving power well above 11000 in both $hv_{in}$ and $hv_{out}$ near a photon energy of 930 eV, with a vast potential for improvement. Combining such a spectrometer – nicknamed $hv^2$ – with an XFEL source allows efficient time-resolved RIXS experiments.

**Keywords: Resonant inelastic X-ray scattering, X-ray optics, X-ray spectrometers, free-electron lasers**

1. Introduction

High-resolution RIXS (Resonant Inelastic X-ray Scattering) is a synchrotron radiation based photon-in / photon-out experimental technique, which gives information about charge-neutral low-energy excitations in correlated electron systems (e.g. crystal field, charge transfer or spin excitations) in solids, liquids and gases (Kotani & Shin 2001).

Detalization of the physical information available from RIXS is directly connected to the resolution and detection efficiency of the RIXS instrumentation (for some entries see Nordgren *et al*. 1989; Hague *et al*. 2005; Tokushima *et al*. 2006; Ghiringhelli *et al*. 2006). Soft-X-ray RIXS spectrometers of the last generation (Ghiringhelli *et al*. 2006) based on variable line spacing (VLS) gratings and high-resolution CCD detectors allow routine operation with a resolving power $E/\Delta E$ better than 10000 at 1 keV photon energy, which takes the RIXS experiment from the energy scale of charge-transfer and crystal-field excitations to that of orbital and magnetic excitations (Schlappa *et al*. 2009). The position-sensitive CCD detector allows parallel detection in outgoing photon energies $h\nu_{out}$. Modern beamlines, delivering the spot size on the sample in a few μm range, enable slitless operation of the RIXS spectrometers, dramatically increasing their detection efficiency. Interesting is a non-conventional concept of active grating monochromator – active grating spectrometer (Fung *et al*. 2004) which promises an increase of the detection efficiency by 2 orders of magnitude.

The full set of RIXS data is a two-dimensional map of X-ray scattered intensity $I(h\nu_{in},h\nu_{out})$ depending on the incoming $h\nu_{in}$ and outgoing $h\nu_{out}$ photon energies. Presently, one acquires $I(h\nu_{in},h\nu_{out})$ in a sequental fashion by measuring one-dimensional $I(h\nu_{out})$ spectra over a series of separate $h\nu_{in}$ selected by the beamline monochromator. Here, a concept is presented of the RIXS spectrometer to enable acquisition of the whole $I(h\nu_{in},h\nu_{out})$ map in one shot with parallel detection in $h\nu_{in}$ and $h\nu_{out}$.

## 2. Concept

Optical scheme of such a RIXS spectrometer – nicknamed $h\nu^2$ for simultaneous detection in $h\nu_{in}$ and $h\nu_{out}$ – is shown in Fig. 1. The monochromator produces in its (stigmatic) focal plane a line image of light with vertical dispersion in energy $h\nu_{in}$. A refocusing KB optics consisting of the vertically and horizontally refocusing (plane-elliptical) mirrors *V-RM* and *H-RM* brings this image into a line focus on the sample. Extreme vertical demagnification of the refocusing stage is necessary to squeeze the vertical extension of the image, important for inhomogeneous samples, and extreme horizontal demagnification to deliver small source size and thus high energy resolution for the spectrometer operated slitless. The light scattered from the sample is intercepted by a vertically focusing (plane-elliptical) mirror *FM* of the spectrometer stage. It operates in the vertical plane to bring the scattered light image into a magnified image on the two-dimensional position-sensitive detector (PSD) with vertical dispersion in $hv_{in}$. A (spherical) VLS *grating* operates in the horizontal plane to disperse the scattered light in $hv_{out}$ and focus it onto the PSD with horizontal dispersion in $hv_{out}$. In this way the *full two-dimensional image* of RIXS intensity $I(hv_{in}, hv_{out})$ is formed on the PSD in one shot with parallel detection in $hv_{in}$ and $hv_{out}$.

A few comments:

• If one inserts a slit in the monochromator focal plane to select one single $hv_{in}$, the $hv^2$ spectrometer becomes fully equivalent to standard RIXS spectrometers with VLS gratings (Ghiringhelli *et al.* 2006) where the dispersive plane is horizontal and the H-FM acts to focus the beam on the CCD in the non-dispersive vertical plane, increasing the spectrometer acceptance by a factor of ~3 compared to the single-grating design (Hague *et al.* 2006). Therefore, the $hv^2$ concept adds simultaneous detection in $hv_{in}$ as a 'free lunch' *without compromising the detection efficiency* in $hv_{out}$ compared to the standard RIXS spectrometers. There is no compromise on resolution either provided the horizontal demagnification is sufficient;

• The $hv^2$ spectrometer inherently includes an option for *XAS data acquisition* in the total fluorescence yield (TFY) in one shot of parallel detection in $hv_{in}$. This measurement mode is realized simply by setting the spherical grating to the zero diffraction order. The vertical line formed in this case on the PSD is the XAS spectrum as a function of $hv_{in}$. Note that aberrations from the grating affect in this case only the horizontal image profile and, by integration in the horizontal direction, have no influence on the XAS spectrum. In this connection one should mention the Energy-Dispersive XAS (see a review in Pascarelli *et al.* 2006) which is an efficient technique for one-shot XAS data acquisition with hard X-rays. This techniques uses however transmission of X-rays through the sample and is not applicable in the soft-X-ray range;

- Response of the $hv^2$ spectrometer in $hv_{in}$ critically depends on homogeneity of the sample along the line focus. However, the vertical refocusing can reduce its extension below 100 μm within an $hv_{in}$ bandwidth of a few eV (see the ray tracing analysis below) which is acceptable for most of the samples. The sample homogeneity is not an issue for studies on liquids and gases;

- An important aspect of the $hv^2$ concept compared to the usual sequential data acquisition is that the full RIXS snapshot $I(hv_{in},hv_{out})$ of the electronic structure is acquired in the *same instant of time*. This is a crucial advantage for studies of time evolution processes, including kinetics of chemical reactions;

- The total optical length of the spectrometer stage has to be constant under variation of photon energy in order to stay focused in $hv_{in}$. Maintaining high resolution in $hv_{out}$ requires then two degrees of freedom of the optical system to cancel both the defocus and coma aberrations (Strocov *et al.* 2008). They can be delivered by (coordinated) variation of the grating pitch angle and the grating translation.

The $hv^2$ concept promises adequate performance already with common 3$^{rd}$ generation synchrotron sources (see below). However, its full potential unfolds with X-ray free-electron laser (XFEL) sources. Their inherently pulsed operation ideally combines with the $hv^2$ spectrometer ability to produce full $I(hv_{in}, hv_{out})$ snapshots in the same instant of time, allowing efficient time-resolved measurements (Patterson *et al*. 2009). Furtermore, the second data dimension radically alleviates the problem of low average intensity of the XFEL radiation. On the technical side, an important advantage of XFELs is a *round spot profile*. Much smaller horizontal source size compared to the synchrotron sources allows further reduction of the horizontal spot size on the sample and thus increase of resolution in $hv_{out}$. Furthermore, small angular divergence of XFEL radiation helps reduction of the size of the optical elements.

## 3. Ray-tracing simulations

Expected performance of the $hv^2$ spectrometer installed at a typical undulator beamline of a 3$^{rd}$ generation synchrotron source was estimated with preliminary ray tracing simulations. No attempt ws made to optimize exact parameters of the optical scheme. The simulations were peformed with the code PHASE (Bahrdt *et al*. 1995). In order to reduce influence of the polynomial approximation of the ellipsoidal shape (Peatman 1997) in this code, the ray tracing was restricted around the central ray within an angular divergence of 0.01 mrad for the refocusing stage and 0.1 mrad for the spectrometer. The refocusing and spectrometer stages each had a length of 5500 mm. The light incidence angle was 88$^o$ for all optical elements. The slope errors of all optical elements were taken as 0.5 µrad, which is a conservative state-of-art value for the (bendable) plane optics. The energy was chosen as 930 eV (near the $2p_{3/2}$ core level of Cu).

*Monochromator* adopts the classical collimated-light PGM scheme (Follath & Senf 1997) with stigmatic focus. The parameters of the source and positions of the optical elements are chosen close to the ADRESS beamline of SLS (Strocov *et al*. 2009). In order to increase the horizontal demagnification, the collimating mirror is chosen cylindric, with the horizontal focusing performed by the focusing mirror having the toroidal shape. With a grating of 1400 lines/mm and $C_{ff} = 2.75$, the monochromator delivers $E/\Delta E \sim 15000$. The image in its focal plane formed by 5 monochromatic rays separated by 1 eV is shown on Fig. 2 (*left*).

*Refocusing stage* has a total length of 5500 mm. The V-RM is installed in 500 mm from the sample and H-RM in 200 mm, delivering extreme horizontal demagnification of the spot (combining the two KB mirrors in one ellipsoidal mirror results in a 'smiley' distortion of the image on the sample). With the light incident on the sample at a grazing angle of 30° and scattering angle 90°, the image seen by the spectrometer is shown in Fig. 2 (*center*). Its vertical extension is less than 100 μm. Note certain distortion of the vertical scale due to spatial extension of our polychromatic source. Each spot has a vert. × hor. FWHM size of $1.7 \times 9.2$ μm$^2$.

*Spectrometer stage* has the V-FM in 200 mm from the sample, delivering extreme vertical magnification of the image on the PSD to reduce influence of its pixel size. The optical element dispersing the light in $h\nu_{out}$ is a spherical VLS grating with a central groove density $a_0$ of 3500 lines/mm installed in ~1500 mm from the sample (optimization of the grating position and VLS parameters followed the procedure from Ghiringhelli *et al*. 2006). The PSD is turned to a grazing incidence angle of 20° in the horizontal plane to reduce the effective pixel size for $h\nu_{out}$. Certainly the spectrometer stage can use other optical schemes such as the Hettrick-Underwood one (Hague *et al*. 2006) bringing larger acceptance in the dispersive plane.

The image formed on the PSD is shown in Fig. 2 (*center*). The monochromatic lines are now dispersed in $h\nu_{out}$ (= $h\nu_{in}$ in our case) in the horizontal direction. This image is therefore the *full map of scattered intensity* resolved in $h\nu_{in}$ and $h\nu_{out}$. In convolution with an effective PSD spatial resolution of 24 μm (typical of the nowadays CCD detectors) the image demonstrates $E/\Delta E$ in $h\nu_{in}$ and $h\nu_{out}$ better than 11100 and 13700, respectively, allowing state-of-art RIXS experiments. Certain distortion of the $h\nu_{in}$ scale can easily be corrected by post-processing of the data.

The above preliminary optical scheme has a vast room for improvements. The resolution in $hv_{in}$ can be increased by optimization of the refocusing stage vertical demagnification, which defines the vertical image size on the sample and thus on the PSD. Most important, one can implement the monochromator without any horizontal focusing to deliver divergent light *directly* to the H-RM. The concomitant increase in the nominal horizontal demagnification by a factor ~3 will dramatically reduce the horizontal spot size on the sample and thus improve the spectrometer resolution in $hv_{out}$.

## 4. Summary

A concept of spectrometer for resonant inelastic X-ray scattering (RIXS) – nicknamed $hv^2$ – has been presented where imaging and dispersion actions in the vertical and horizontal planes, respectively, are combined to deliver full two-dimensional map of RIXS intensity in one shot with parallel detection in $hv_{in}$ and $hv_{out}$. This scheme is free of any compromise on energy resolution or detection efficiency compared to the conventional RIXS spectrometers. Preliminary ray-tracing simulations with a 3$^{rd}$ generation synchrotron source demonstrate a resolving power above 11000 in $hv_{in}$ and above 13700 in $hv_{out}$ at 930 eV photon energy, with a vast room for further improvements such as a monochromator without intermediate horizontal focusing. Combining the $hv^2$ spectrometer with an XFEL source allows efficient time-resolved RIXS experiments with further increase of resolution.

*Acknowledgments.* The author thanks U. Flechsig for expert advices, and F. van der Veen, C. Quitmann, B. Patterson, Th. Schmitt and L. Patthey for encouragement and promoting discussions.

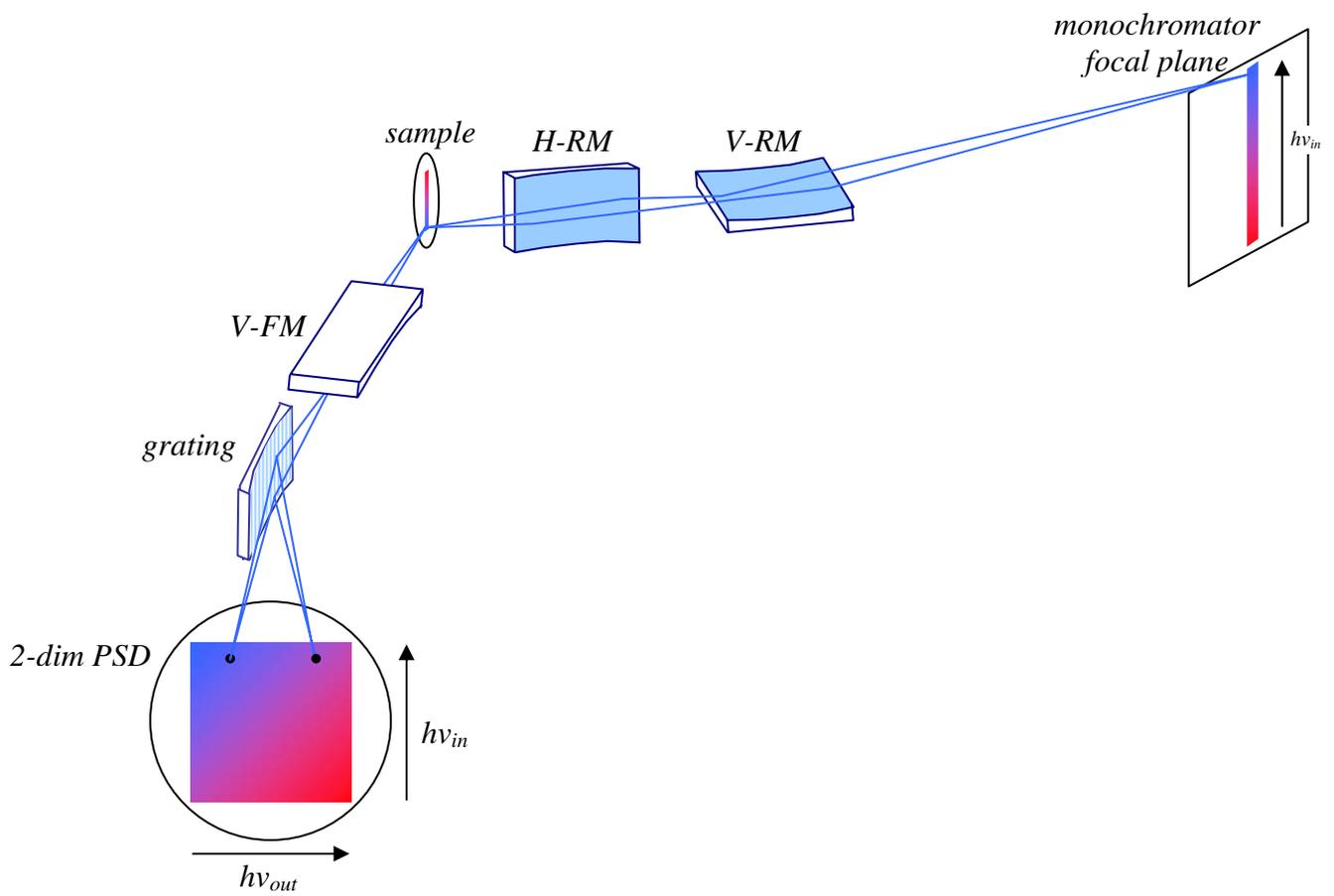

Fig. 1. Operational principle of the $hv^2$ spectrometer. The image formed on the PSD is the *full two-dimensional image* of RIXS intensity $I(hv_{in}, hv_{out})$ acquired with parallel detection in $hv_{in}$ and $hv_{out}$.

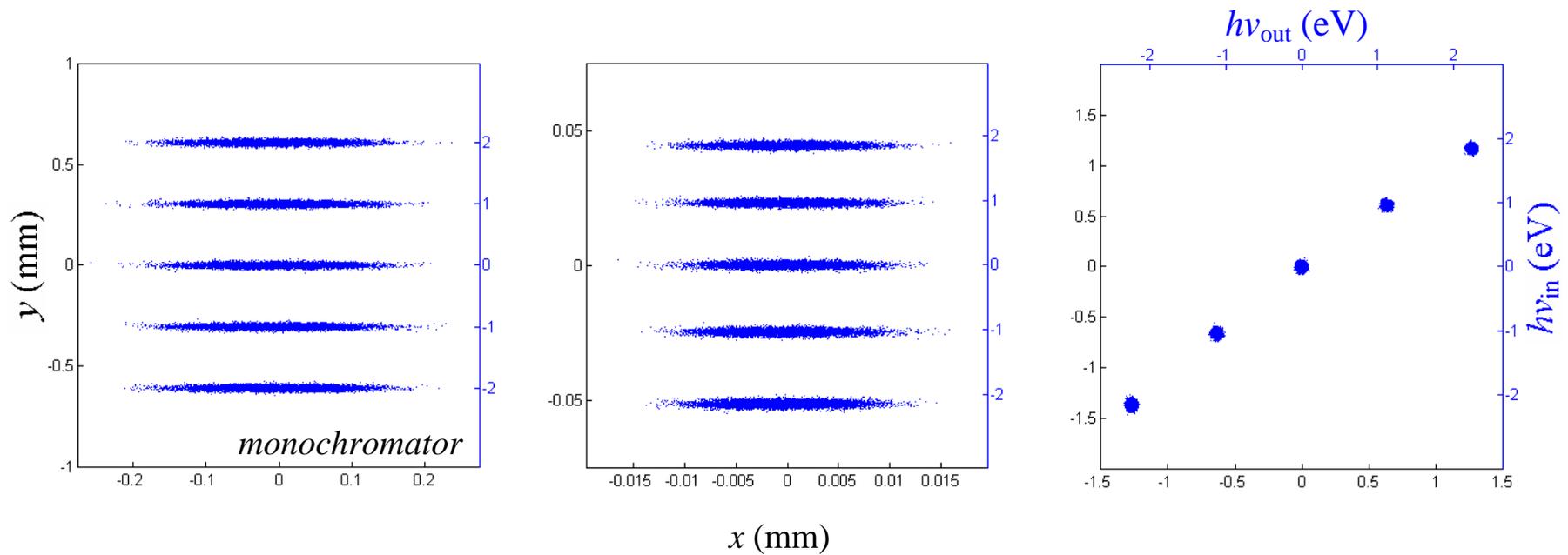

Fig. 2. Ray-tracing simulations of images formed by 5 monochromatic rays separated by 1 eV: (*left*) in the monochromator focal plane, (*center*) on the sample in scattered intensity seen by the spectrometer, and (*right*) on the two-dimensional PSD resolved in $h\nu_{in}$ and $h\nu_{out}$. Note different coordinate scales and the same $h\nu_{in}$ energy scale.